\documentclass[referee]{aa} 
\usepackage{graphics}

\begin{document}


   \title{Molecular dynamics simulation of gaseous atomic hydrogen interaction with hydrocarbon grains}
\author{R. Papoular
\inst{1}
          }

   \offprints{R. Papoular}

   \institute{Service d'Astrophysique and Service de Chimie Moleculaire, CEA Saclay, 91191 Gif-s-Yvette, France\\
              e-mail: papoular@wanadoo.fr
             }

   \date{ }
   \authorrunning {R. Papoular}
   \titlerunning {Gas-surface interactions}
   \maketitle
\begin{abstract}

Semi-empirical molecular dynamics is used to simulate several gaseous atomic hydrogen interactions with hydrocarbon grains in space: recoil, adsorption, diffusion, chemisorption and recombination into molecular hydrogen. Their probabilities are determined as a function of initial velocity of gas atoms. The equilibrium hydrogen coverage of free grains is then derived. These data can be used in calculations of material and energy balance as well as rates of hydrogen recombination on grains. The role of grain temperature is also considered.
 %
  \end{abstract}

%
\section{Introduction}

The primary purpose of the present work is to study several processes which occur when H atoms impinge on carbonaceous grains, and thus bear upon the recombination of these atoms into diatomic molecules. This, in turn, affects chemical and energetic equilibria in gas and grains. Here, focus is set on such equilibria in environments typical of carbon-rich PDRs (Photon-Dominated Regions). Gas-surface interaction has been the subject of intensive and extensive experimental work (e.g. Winkler and Rendulik \cite{wink92}, Donnet et al. \cite{don93}) as well as an abundant literature, including several treatises (see  Bortolani et al. \cite{bort94}, Masel \cite{mas96}). In general, however, available information relates to very low target temperatures (e.g. Pirronello et al. \cite{pir99}, Katz et al. \cite{katz99}, Cazeaux and Tielens \cite{caz04}) and/or pure and regular dielectric or metallic surfaces of technological interest (see many issues of Surface Science). Here, instead, we are interested in hydrogenated, amorphous, semiconducting carbon surfaces. 

When an H atom hits such a surface, it may \emph{recoil} and be scattered away. Or it may be attracted and captured by a dangling C-bond (dehydrogenated site) and stick to the surface (strong chemisorption). Otherwise, it may interact electronically with one particular H atom of the surface, so strongly as to form an H$_{2}$ molecule, which promptly leaves the surface. This process is called \emph {abstraction} or E-R (Eley-Rideal) process; it can be \emph{direct} or \emph{indirect} (retarded) according to whether it takes place promptly or after some wandering over the surface (see  Stecher and Williams \cite{stec66}). Such surface \emph{diffusion} is induced by the initial velocity of the projectile and enhanced by thermal agitation of the target atoms.

If the H projectile has a low initial velocity, it may be \emph{adsorbed} in a shallow (physisorption) well, making a weak bond to the surface, which allows diffusion. When the grain temperature is not too high, the atom may stay on the surface for a long time. Then, a second projectile may follow a similar course and diffuse to meet the first one, in which case a molecule is formed and leaves the surface: this is the H-L (Hinshelwood-Langmuir) process. 
Both the E-R and H-L mechanisms contribute to the \emph{recombination rate} R (cm$^{3}$s$^{-1}$; see Duley and Williams \cite{dul84}), in various amounts, according to gas density and grain temperature.

Finally, if the projectile velocity is very high, it may kick out the target H atom (\emph{expulsion}) and, sometimes, substitute to it at the same C site (\emph{substitution}). The \emph{hydrogen coverage} of the grain surface, and hence the recombination rate as well as the IR (InfraRed) spectrum, strongly depend on the balance of all these processes, which may all occur on the same grain when the gas atom velocities cover a wide range.

All these processes are sensitive to velocity, angle and site of impact; hence the great difficulty of the experimental approach. As to the classical theoretical approach, it requires tailoring essential parameters (e.g. activation energies; see Duley \cite{dul96}, Habart et al. \cite{hab04}) to fit observations. Parneix and Brechignac (\cite{parn98}), instead, used numerical simulation to study H$_{2}$ recombination on graphite \emph{from first principles}. Here, Molecular Dynamics is also used to simulate numerically not one particular, but all the above phenomena occuring simultaneously. For this purpose, the target is not chosen to be a surface, but a finite, small grain, and not one, but several projectiles are launched at the same time towards it from different directions, with the same velocity, $V_{0}$, set, at the start of each run, at a different value, between 1 and 500 km.s$_{-1}$ (which includes the range of gas temperatures of interest, between 0 and 1000 K). Also, because the grain temperature is much higher in PDRs than in molecular clouds, the model target was given initial temperatures ranging from 0 to 365 K. One essential feature of this simulation is the fact that \emph{the target structure is not rigid}: both the nuclear and electronic positions change under the electrical influence of the incoming H atom, allowing energy exchange between them; otherwise, only elastic recoil is possible. More generally, all the physics involved is included a priory in the chemical code which governs the dynamics, and no assumption is made on any of the parameters of interest (e.g. sticking coefficient on, or H coverage of, grains, or H$_{2}$ photo-dissociation rate or activation energies).

\section{The chemical code}
A description of the software package used for simulation can be found in Papoular (\cite{pap01}). The particular code used here is AM1 (Austin Model 1), as proposed, and optimized for carbon-rich molecules, by Dewar et al. (\cite{dew85}). This semi-empirical method combines a rigorous quantum-mechanical formalism with empirical parameters obtained from comparison with experimental results. It computes approximate solutions of Schroedinger's equation, and uses experimental data only when the Q.M. calculations are too difficult or too lengthy. This makes it more accurate than poor ab initio methods, and faster than any of the latter.

The molecular dynamics relies on the Born-Oppenheimer approximation to determine the motions of atoms under nuclear and electronic forces due to their environment. At every step, all the system parameters are memorized as snapshots so that, after completion of the run, a movie of the reactions can be viewed on the screen, and the trajectory of any atom followed all along. This makes for a better understanding of the details of  mechanisms and outcomes. Note that the dynamics of bond dissociations and formation can only be simulated by using Unrestricted Hartree-Fock (UHF) wave functions in the Q.M. part of the calculation (see Szabo and Ostlund \cite{sza89}). The elementary calculation step was set at $10^{-15}$ s for initial projectile velocities smaller than 200 km.s$^{-1}$ and $10^{-16}$ above.

Semi-empirical simulation methods do not account for purely quantum-mechanical phenomena like tunneling or zero-point energy. In the present context of relatively high grain temperature, the former is negligible as regards surface diffusion, and the latter hardly affects the probability and energetics of the physico-chemical reactions involved.

\section{The model dust grain}

This study is restricted to carbonaceous IS (InterStellar) dust. IR spectra of such dust reveals the presence of both aromatic and aliphatic C-C and C-H bonds (e.g. Pendleton et al. \cite{pend94}). Graphitic dust has been abundantly modeled (see Parneix and Brechignac \cite{parn98}, Ko et al. \cite{ko02}). Here, the focus is on amorphous aliphatic dust. In order to provide a model of fully hydrogenated surface (Fig. 1, right), one 3-D structure of 100 atoms was assembled from 66 C atoms and 34 H atoms, forming two 6-membered, one 5-membered and one 4-membered rings, and several aliphatic chains. 11 H atoms are involved in CH sites or functionalities (tertiary carbon), 44 in CH$_{2}$ sites (secondary carbon) and 9 in CH$_{3}$ sites (primary carbon). The bond lengths and angles of this initial structure were then automatically optimized by the software, searching for the minimum overall potential energy. The final system is enclosable in a box of size 15*12*7 AA{\ }$^{3}$ (the cluster sizes adopted here are a compromise between computational time and relevance to real IS dust). In a preliminary run, the system is brought up to the desired temperature, T, meaning that the atoms are given random velocities, increasing progressively in a predetermined number of calculation steps, subject to bond constraints, up to a Maxwellian distribution characteristic of T. Such a set-up provides information on all the processes described above, except recombination of H atoms on dangling C-bonds, which are absent here. The latter is better studied by means of the second, completely de-hydrogenated, target, Fig. 1, left. It includes 33 C atoms, of which 2 C-sites with 3 dangling bonds, 10 sites with 2, 11 with only 1 and 10 with no such bond. Its size is 9*5*4 \AA{\ }$^{3}$.

In order to study the energy balance, it is necessary to know the net result of each elementary process, which requires knowledge of the C-H and H-H bonding energies. These are obtained from the cold, optimized targets by extracting a particular H atom and noting the increase in overall potential energy. They are thus found to be 109.4 kcal/mole for H$_{2}$, 107 for aromatic C-H and about 96 for all aliphatic C-H bonds. The corresponding values in eV are 4.75, 4.65 and 4.15 respectively, and the bond lengths, 0.68, 1.1 and about 1.1 AA{\ }, respectively. By contrast, a \emph{weak} chemical bond, 2.1 eV, is formed with one C atom of a graphitic plane by an H atom, 1.15 \AA{\ } above the C atom, which is itself slightly uplifted in the process.

Physisorption sites have been found by minimizing the potential energy of an H atom deposited at some arbitrary, but typical, locations around and at a few angstroms from the target. For the hydrogenated aliphatic grain, most of them are of 2 types: a) at 2.85 \AA{\ } from a C atom, along the C-H bond, the potential well depth or bond energy is 45 meV, same as for an aromatic C-H site studied independently; b) between two adjacent H atoms, at about 2 \AA{\ } from both, the well depth is about 80 meV. For the dehydrogenated grain, physisorption sites are absent due to the long range attraction of the dangling bonds.

In dynamic calculations, no activation barrier for abstraction was found in the exact direction of aliphatic C-H bonds, but there is an increasing potential barrier as the angle of incidence increases, and no abstraction was observed beyond 60 degrees. By contrast with aliphatic bonds and \emph{weak} bonds above a graphitic plane, no abstraction at all was observed in the case of \emph{strong} aromatic C-H bonds. This must be traced to the greater rigidity of the latter. Indeed, \emph{freezing} the C-H bond (i.e. letting it retain the same length and direction during the calculation) suppresses abstraction even in aliphatic sites. The importance of bond elasticity is highlighted by the direct observation of the \emph{harpoon} effect (see Atkins\cite{atk98}) just before abstraction occurs: as the projectile approaches a target H atom, both their electron distributions change so that the attraction increases between them while the adjacent C-H bond expands. These changes are reversible until the \emph{saddle point}, or \emph{transition state}, is reached, when the distance between the H atoms is 1.3 \AA{\ }. Then, this distance is suddenly reduced by one half and the H$_{2}$ molecule is formed and escapes. The saddle point clearly shows up in the potential curve, in the form of a small bump within the main well. No such phenomenon occurs with the aromatic C-H bond.

\begin{figure}
\resizebox{\hsize}{!}{\includegraphics{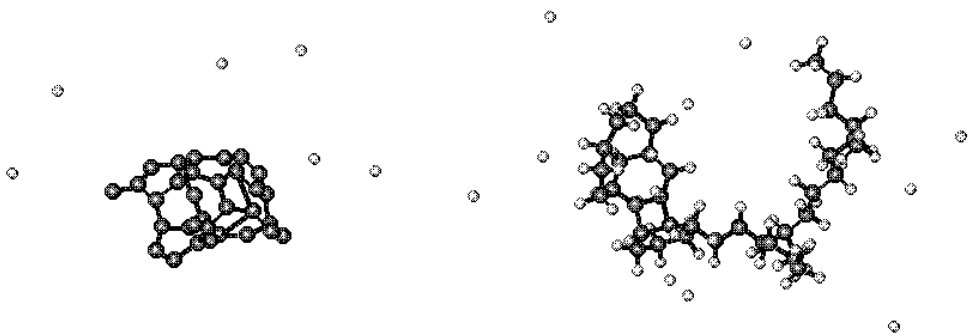}}
\caption[]{Right: the hydrogenated target with saturated C-bonds; 64 C atoms (larger circles) and 36 H atoms (smaller circles); enclosing box 15*12*7 \AA{\ }$^{3}$. Left: the pure-carbon target with dangling (free) bonds; 33 C atoms; 9*5*4 \AA{\ }$^{3}$. The isolated H atoms surrounding the targets are the projectiles, 10 and 6 respectively, which are all given the same initial velocity, directed to a central C atom in each target.} 
\end{figure}

\begin{figure}
\resizebox{\hsize}{!}{\includegraphics{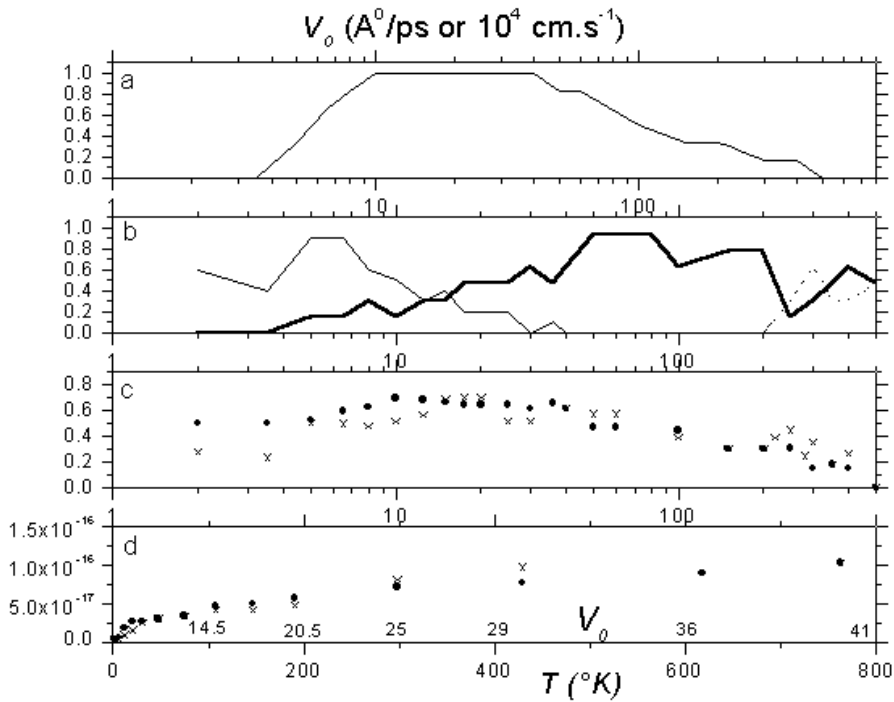}}
\caption[]{Figure 2a: recombination probability coefficent,$k_{c}$, for the pure -C target at 0 K initial temperature. Figure 2b: probability coefficients for the H-saturated target at 0 K: prompt or direct abstraction  ($k_{a}$, heavy full line); transient physisorption or indirect abstraction ($k_{p}$, thin); expulsion ($k_{e}$, dots). Figure 2c: Hydrogen coverage,$c_{H}$, of a grain in equilibrium at 0 K (dots) and 350 K (crosses). Figure 2d: the recombination rate, $R$, on a grain in equilibrium at 0 K (dots) and 350 K (crosses). All $k_{r}$'s, and hence coverage and $R$, are functions of the initial velocity, $V_{0}$, assumed the same for all projectiles. Note the different scale for fig. 2d.}
\end{figure}

\section{Data analysis and results}

In a typical experiment, the target is surrounded by 10 or 6 H atoms (for the hydrogenated and dehydrogenated targets, respectively) at a distance larger than 4 \AA{\ } and as far as possible from each other, to minimize interference between them and occurences of L-H events between adsorbed projectiles. Each is given a velocity $V_{0}$, directed to an arbitrary, but central, atom of the target; V$_{0}$ spans the range between 1 and 500 \AA{\ } ps$^{-1}$ ($10^{4}$ to $5 \,10^{6}$ cm.s$^{-1}$) corresponding to impact energies from $5 \,10^{-5}$ to 12.5 eV. The calculation is then started and the trajectory of each projectile and target atom is followed in time and visualized on the computer screen. The calculation extends as long as necessary for all projectiles to interact (almost) completely: this usually takes less than 2 ps (2 10$^{-12}$ s). 

Let $N_{r}$ be the number of completed reactions of type $r$ between a projectile and an H or C site of which $s_{r}$ are available out of a total of $s$ target sites. A corresponding coverage is defined as $c_{r}=s_{r}/s$. If a single projectile were used and the experiment repeated a large number of times, $p$, so that $N_{r}$ were very large, a  probability could be defined for reaction $r$ by the first of the two equalities

$p_{r}=N_{r}/p= k_{r}.c_{r}$;

the second equality defines $k_{r}$ as a coefficient specific of this reaction which is, in general, a function of $V_{0}$ and temperature T. Here, computer time is saved by using $p$ projectiles simultaneously in each run instead of repeating the run $p$ times and, in an attempt to capture the main features of the behaviour of $k_{r}$ for the various possible reactions, the crude assumption is made, that the notion of a probability $p_{r}$ may be extended to the present calculations.  Now, that probability is, by definition, proportional to the number of relevant sites and to a corresponding effective crossection; the proportionality constant  is found by rewriting the coverage as $(s_{r}/S).(S/s)$, where $S$ is the surface area of the grain. Whence the effective crossection per such site, $\sigma_{r}=k_{r}.S/s$, which gives a geometrical meaning to $k_{r}$. Since the distance between two adjacent target sites is in the range 1 to 1.6 \AA{\ }, $S/s$ may be estimated at 2 \AA{\ }$^{2}$.

Figure 2a represents the  probability coefficient $k_{c}$ for recombination on the dehydrogenated target (at zero initial temperature), as deduced from the second of the equalities above, with $c_{r}$=1 and p=6. There seems to be a threshold for recombination at about $5 \,10^{4}$ cm.s$^{-1}$, indicating an activation energy of $\sim$35 meV. This, however, is not associated with any bump in the static potential curve. It is due to the fact that the approaching projectile continuously gives away  some of its kinectic energy to the deformable target. If and when it comes to a standstill, at $\sim4$ \AA{\ } from the target, the latter relaxes and repels the projectile. Beyond velocity threshold, the latter retains all along enough energy to reach the point of no return, at the edge of the chemisorption well.

Far beyond threshold, the argument is reversed: recombination is only possible if the projectile can give away enough of its energy to the target. As $V_{0}$ increases, this becomes more and more difficult because the amount of energy to be evacuated increases while the time available to do that shortens. Hence, the recoil probability increases and recombination becomes rarer, while the minimum distance of approach before recoil decreases and tends to $\sim0.5$ \AA{\ }. At intermediate velocities, optimal conditions for recombination are met and its probability reaches 1. The target temperature hardly affects these results: at 350 K, the velocity threshold for recombination is very slightly increased.

Figure 2b represents the probability coefficients for (transient) physisorption on, and prompt (direct) abstraction and expulsion of H atoms from, the saturated target at zero initial temperature. Here, p=10 and $c_{r}$ is 1, 0.64 and 0.64 respectively for the 3 types of reaction. The threshold for expulsion is 250 \AA{\ }/ps and its probability increases quickly beyond. The corresponding threshold energy is $\sim3.1$ eV, as compared with 4 eV bond energy; the difference is contributed by target deformation. 

At velocities lower than 20 \AA{\ }/ps, contrary to the dehydrogenated case, a projectile can now be caught in the physisorption belt, and diffuse all around the target. During this journey, it ultimately abstracts an H atom from a hydrogenated site (delayed or indirect abstraction) or recombines on a dangling C bond, if there is any.  Most often, this occurs in less than 2 ps because diffusion is very fast ($>$1 \AA{\ }/ps). At higher grain temperatures, the projectiles are more easily repelled before hitting the surface: at 350 K, the probability of physisorption is nearly uniformly reduced by a factor 2; also, diffusion is accelerated, but this temperature is not high enough for the adsorbate to be desorbed. In the following, we shall therefore assume that, for gas densities and temperatures of interest, recombination or indirect abstraction happens before another projectile stands a chance of being physisorbed and participating in a H-L recombination process. 

With these coefficients at hand, it is now possible to derive the hydrogen coverage of a free grain. Assuming exact balance between abstraction ($k_{a}$ with a delayed contribution by diffusing adsorbed atoms, $k_{p}$) and  expulsion ($k_{e}$) from hydrogenated sites, on the one hand, and recombination ($k_{c}$ with a delayed contribution by diffusing adsorbed atoms) on the other hand, the equilibrium coverage is found to be $c_{H}=1-c_{C}=\frac{k_{c}+k_{p}}{k_{a}+k_{e}+k_{c}+k_{p}}$. Figure 2c represents this coverage for 0 and 350 K initial temperatures. They differ notably only below 5 \AA{\ }/ps, because of the different probabilities of adsorption. By definition (see above), knowledge of $c_{H}$ and $k_{r}$ finally allows one to determine the probability, $p_{r}$, of each of the surface phenomena of interest, which, in turn, is necessary to determine the energy balance. Thus, the probability of recombination per H atom impact on a typical grain is: $P=(k_{c}+k_{p}) c_{H}$. Now, the recombination rate is $R=1/3 \,V_{0}P\Sigma$, where $\Sigma$ is the (carbonaceous) grain surface per H atom in space and is taken to be $2\,10^{-21}$ cm$^{2}$. This quantity is represented in Fig. 2d, as a function of an effective \emph{gas} temperature $T_{eff}=\pi m_{H}V_{0}^{2}/8k_{B}$. In effect, the $k_{r}$'s should be replaced by integrals over the curves of Fig. 2a and 2b weighted by a distribution of velocities. Figures 2c and 2d show that \emph{grain} temperature in this range has little relevance to H coverage or $R$. On the other hand, $R$ is strongly dependent on gas temperature and dust abundance. Considering the large uncrtainties on the latter, the trend and range of $R$ are in rough agreement with those deduced from observations by Habart et al. (\cite{hab04}).

 Regarding energy balance, note that, when recombination of an H atom occurs on a target dangling bond, all the available chemical energy (the C-H bond energy) is deposited in the grain, as is the initial projectile energy. By contrast, when abstraction occurs, the available energy is the sum of the projectile kinetic energy and the H-H bond energy but most of it goes into breaking the target C-H bond and feeding the internal and external kinetic energies of the newly formed H$_{2}$ molecule. The detailed break-up depends on the particulars of the impact, but little kinetic energy is left in the grain and the molecule leaves promptly with a velocity higher than 70 \AA{\ }/ps. Similarly, in recoil, only a small fraction of the projectile kinetic energy is deposited in the target.

\end{document}